\documentclass{ws-procs961x669}            
\usepackage{physics}
\begin{document}
\title{Tractor beams, pressor beams, and stressor beams\\
 within the context of general relativity}
\author{Matt Visser and Jessica Santiago}
\address{$^1$ School of Mathematics and Statistics, Victoria University of Wellington, \\
PO Box 600, Wellington 6140, New Zealand.\\
E-mail: matt.visser@sms.vuw.ac.nz\\
E-mail: jessica.santiago@sms.vuw.ac.nz}
\author{Sebastian Schuster}
\address{Institute of Theoretical Physics,
Faculty of Mathematics and Physics, Charles University,\\
V~Hole\v{s}ovi\v{c}k\'{a}ch~2,
180 00 Prague 8,
Czech Republic\\
E-mail: sebastian.schuster@utf.mff.cuni.cz}

\begin{abstract}
Both traversable wormholes and warp drives, concepts originally developed within the context of science fiction, 
have now (for some 30 odd years) been studied, debated, and carefully analyzed within the framework of general relativity. An overarching theme of the general relativistic analysis is unavoidable violations of the classical point-wise energy conditions. 
Another science fiction trope, now over 80 years old, is the tractor beam and/or pressor beam. 
We shall discuss how to formulate both tractor beams and/or pressor beams, and a variant to be called a stressor beam, within the context of reverse engineering the spacetime metric.  
(While such reverse engineering is certainly well beyond our civilization's current capabilities, we shall be more interested in asking what an arbitrarily advanced civilization might be able to accomplish.) 
We shall see that  tractor beams and/or pressor beams can be formulated by suitably modifying the notion of warp drives, and that, as for wormholes and warp drives, violations of the classical point-wise energy conditions are utterly unavoidable.
\end{abstract}


\bodymatter
\def\d{{\mathrm{d}}}
\def\L{{\mathcal{L}}}
\parindent0pt
\parskip7pt

\section{Introduction}

Traversable wormholes have now been extensively studied for some 33 years.~\cite{Morris:1988a, Morris:1988b, Visser:1989a, Visser:1989b, Roman:1992, Cramer:1994, Kar:1994, Kar:1995, Visser:1995, Ford:1995, Teo:1998, Hayward:1998, Visser:2003, Lemos:2003, Roman:2004, Kar:2004, Muller:2008, Hayward:2009, James:2015}  The general theoretical framework has been to precisely define, and then reverse engineer, a physically interesting spacetime metric, using the Einstein equations to determine what matter sources some arbitrarily advanced civilization might need to employ in order to generate such interesting spacetime geometries.   Perhaps the key feature of reverse engineering is that adopting this point of view reduces the problem of solving the Einstein equations to a relatively simple exercise in differentiation, instead of the much harder problem of integrating the underlying nonlinear PDEs. 

A key  feature of traversable wormhole spacetimes, at least within the usual context of standard general relativity, is the utterly unavoidable violation of the classical energy conditions.~\cite{Morris:1988a, Morris:1988b, Visser:1989a, Visser:1989b, Roman:1992, Cramer:1994, Kar:1994, Kar:1995, Visser:1995, Ford:1995, Teo:1998, Hayward:1998, Visser:2003, Lemos:2003, Roman:2004, Kar:2004, Muller:2008, Hayward:2009, James:2015} 

Even if one moves beyond the framework of standard general relativity there are unavoidable purely geometric constraints, violations of both the null convergence condition and also violations of the timelike convergence condition, that place very strong constraints on any ``effective'' stress-energy tensor that one might use as a source to generate the spacetime.~\cite{Visser:1995}

Similarly, ``warp drives'' have now been extensively studied for some 27 years.~\cite{Alcubierre:1994, Low:1998, Clark:1999, Lobo:2002, Lobo:2004a, Lobo:2004b, Lobo:2007, Finazzi:2009, Tippett:2013, Alcubierre:2017-book, Alcubierre:2017-basics, Santiago:2021-warp} As is the case for traversable wormholes once one defines a suitable class of warp field spacetimes it is easy to see that reverse engineering the spacetime metric, 
at least within the context of standard general relativity, quite unavoidably leads to significant violations of the classical energy conditions.~\cite{Alcubierre:1994, Low:1998, Clark:1999, Lobo:2002, Lobo:2004a, Lobo:2004b, Lobo:2007, Finazzi:2009, Tippett:2013, Alcubierre:2017-book, Alcubierre:2017-basics,Santiago:2021-warp} Again,  even when stepping outside the framework of standard general relativity there are still unavoidable but purely geometric constraints, violations of the null convergence condition and violations of the timelike convergence condition, that place very strong constraints on any ``effective'' stress-energy tensor.~\cite{Alcubierre:1994, Low:1998, Clark:1999, Lobo:2002, Lobo:2004a, Lobo:2004b, Lobo:2007, Finazzi:2009, Tippett:2013, Alcubierre:2017-book, Alcubierre:2017-basics, Santiago:2021-warp}
(There have recently been several misguided counterclaims regarding these issues which we shall address more fully below, see references~\refcite{Lentz:2020,Bobrick:2021,Fell:2021}. For somewhat related but distinct analyses, with somewhat different flaws, see also the discussion in references~\refcite{Santos-Pereira:2020,Santos-Pereira:2021-a,Santos-Pereira:2021-b,Santos-Pereira:2021-c}, and in reference~\refcite{Beatrix-Drouhet:2020}.)

Much more recently, the last 6 months have seen the current authors  reformulate yet another 
science fiction trope, tractor beams, pressor beams, and stressor beams, within the framework of general relativity.~\cite{Santiago:2021-tractor}
Again we resort to reverse engineering,  and again the classical energy conditions are violated.~\cite{Santiago:2021-tractor}
Even if one steps outside the framework of standard general relativity  there are still purely geometric constraints, violations of the null convergence condition, and also violations of the timelike convergence condition, that make the underlying physics ``exotic'' in some purely technical sense.~\cite{Santiago:2021-tractor}

Of course, in all of these cases one must first precisely define what one means by a traversable wormhole, or a warp drive, or a tractor/pressor/stressor beam. One also needs to carefully assess the extent to  which the energy conditions (or more geometrically, the null and timelike convergence conditions) are to be taken as truly fundamental physics.~\cite{Visser:1995, Santiago:2021-warp, Santiago:2021-tractor, Barcelo:2002}

\section{Energy conditions ---generalities}

It should be emphasized that the energy conditions are not fundamental physics.~\cite{Barcelo:2002} The energy conditions are instead warning flags: They tell you ``here be unusual physics''.
Violations of the energy conditions are not an absolute prohibition;
but they are an indication that one should think very carefully about the underlying physics. 

\clearpage
For instance:
\begin{itemize}
\item 
Observational cosmology (either in the current epoch of the accelerating universe, or during the epoch of cosmological inflation) violates the strong energy condition (SEC).~\cite{Visser:1997,Visser:1999}
\item
Speculative theoretical cosmologies (asymptotically AdS spacetimes) with a negative cosmological constant violate the weak energy condition (WEC).
\item 
Gravitational vacuum polarization (in the test field limit) can violate all the energy conditions, null, weak, strong, and dominant (NEC, WEC, SEC, and DEC).~\cite{gvp1,gvp2,gvp3,gvp4,gvp:mg8}
\item 
Hawking radiation violates all the energy conditions (NEC, WEC, SEC, and DEC).~\cite{Visser:1995}
\item 
Quantum scale anomalies (trace anomalies) violate even the averaged null energy condition (ANEC).~\cite{Visser:1994-scale}
\end{itemize}
But these known energy condition violating scenarios do not necessarily imply that it is possible to concentrate energy condition violating physics in a sufficiently small region to enable ``interesting'' effects --- such as wormholes, warp drives, and the tractor/pressor/stressor beams.\cite{Visser:1995,Visser:2003,Kar:2004}
However, in view of these known energy condition violating scenarios, neither would the necessity of energy condition violations necessarily prohibit  ``interesting'' physics.~\cite{Baccetti:2012re, Martin-Moruno:2013wfa, Martin-Moruno:2013sfa, Hochberg:1998vm, Martin-Moruno:2017exc, Visser:1999fe, Martin-Moruno:2015ena, Morris:1990, Hawking:1991, Kim:1991, Echeverria:1991, Visser:1992a, Visser:1992b, Kay:1996, Krasnikov:1998, Visser:2002, Friedman:1993, Jacobson:1994, Galloway:1999, Friedman:2006}

\noindent
We emphasise that to verify that the energy conditions are \emph{satisfied}:
\begin{itemize}
\item One needs to check \emph{all} relevant ``observers'' --- both \emph{all} physical timelike observers, and in the null limit,  \emph{all} null tangent vectors.
\item One needs at least \emph{some} information concerning \emph{all} of  the stress-energy components.
\end{itemize}
In contrast, to demonstrate the energy conditions are \emph{violated}:
\begin{itemize}
\item It is sufficient to check that \emph{one} timelike observer, (or \emph{one} null curve),
violates the energy conditions.
\item It is sufficient to check that  \emph{some} combination of stress-energy components
violates the energy conditions.
\end{itemize}

\section{Generic warp metric and tractor/pressor/stressor metric}

A sufficiently generic spacetime metric, (suitable for investigating both warp drives and tractor/pressor/stressor beams), 
is to consider:~\cite{Alcubierre:1994, Low:1998, Clark:1999, Lobo:2002, Lobo:2004a, Lobo:2004b, Lobo:2007, Finazzi:2009, Tippett:2013, Alcubierre:2017-book, Alcubierre:2017-basics, Santiago:2021-warp}
\begin{equation}
\d s^2 = - \d t^2  + \delta_{ij}\; \left(\d x^i - v^i(t,\vec x) \,\d t\right) \; 
\left(\d x^j  - v^j(t,\vec x) \,\d t\right). 
\end{equation}

\clearpage
This class of metrics has the properties that it is: 
\begin{itemize}
\item Spatially flat.
\item Unit lapse.
\item The gradients of the flow vector $v^i(t,\vec x)$ must satisfy suitable asymptotic fall-off conditions. 
\begin{itemize}
\item (Otherwise you do not have a localized warp-field.)
\item (You do not want to accelerate the entire universe.)
\end{itemize}
In particular, the spacetime must be asymptotically flat.

\item Three special cases are commonly considered:
\begin{itemize}
\item  Alcubierre: $\vec v(t,\vec x)  = v(t,z) \; \hat z$.
\item Zero-expansion: $\div  \vec v(t,\vec x)  = 0$.
\item Zero-vorticity: $\curl \vec v(t,\vec x)  = 0$.
\end{itemize}
\item
For  tractor/pressor/stressor beam spacetimes one will add extra conditions, such as cylindrical symmetry, and simplify the form of the flow vector $\vec v(t,\vec x) $ by introducing suitable ``envelope'' and ``profile'' functions.
\item
Further generalizations are in principle possible,  but very quickly become calculationally intractable.
\end{itemize}
To start specific computations it is convenient to first pick a co-tetrad:~\cite{Santiago:2021-warp,Santiago:2021-tractor}
\begin{itemize}
\item 
Take the timelike leg to be:
\begin{equation}
(e^{\hat 0})_a = (1;0,0,0)_a =n_a, 
\end{equation}
\item
Take the spatial co-triad to be:
\begin{equation}
\hspace{-25pt}
(e^{\hat 1})_a = (-v_x;1,0,0)_a, 
\quad (e^{\hat 2})_a = (-v_y;0,1,0)_a, 
\quad  (e^{\hat3})_a = (-v_z;0,0,1)_a. 
\end{equation}
\end{itemize}
This \emph{choice} serves to keep computations more-or-less tractable. 

\enlargethispage{30pt}
The corresponding tetrad is then fixed:~\cite{Santiago:2021-warp,Santiago:2021-tractor}
\begin{itemize}
\item 
The timelike leg is:
\begin{equation}
(e_{\hat 0})^a = (1;v^x,v^y,v^z)^a= n^a, 
\end{equation}
\item
The spatial triad is:
\begin{equation}
(e_{\hat1})^a = (0;1,0,0)^a, \quad (e_{\hat2})^a = (0;0,1,0)^a, \quad 
(e_{\hat3})^a = (0;0,0,1)^a. 
\end{equation}
\end{itemize}
With this choice of tetrad the orthonormal components of the extrinsic curvature are now particularly simple:~\cite{Santiago:2021-warp,Santiago:2021-tractor}
\begin{equation}
K_{\hat a \hat b} = \left[ 
\begin{array}{c|c} 0 & 0 \\ \hline 0 & K_{ij}\end{array}\right],
\end{equation}
where
\begin{equation}
K_{ij} = v_{(i,j)}.
\end{equation}

(Effectively one can just quietly drop the hats from covariant spatial components of any tensor in this tetrad basis.~\cite{Santiago:2021-warp,Santiago:2021-tractor})

Brute force computation now yields two easy results, and one somewhat messier result.~\cite{Santiago:2021-warp,Santiago:2021-tractor}
\begin{itemize}
\item 
Easy (Gauss--Coddazzi):
\begin{equation}
\label{E:Gnn}
G_{nn} =  {1\over2} \left(  K^2-\tr(K^2) \right),
\end{equation}
\begin{equation}
\label{E:Gni}
G_{ni} =  K_{ij,j}- K_{,i}.
\end{equation}
\item
Somewhat messier (specialization of ADM formalism):
\begin{equation}
\label{E:Gij}
G_{ij} =  \L_n K_{ij}  + K K_{ij} - 2(K^2)_{ij} -  \left(\L_n K  + {1\over2}K^2 + {1\over2} \tr(K^2)\right) \delta_{ij}.
\end{equation}
\end{itemize}
In order to claim that the energy conditions are satisfied (or claim that the null and timelike convergence conditions are satisfied) one would  need to compute and analyze \emph{all} the components of the Einstein tensor.
(This is one specific place, there are others, at which the claims made in references \refcite{Lentz:2020,Bobrick:2021,Fell:2021} simply fail. See reference~\refcite{Santiago:2021-warp} for an extensive discussion of these issues. The analyses of references~\refcite{Santos-Pereira:2020,Santos-Pereira:2021-a,Santos-Pereira:2021-b,Santos-Pereira:2021-c}, and reference~\refcite{Beatrix-Drouhet:2020} exhibit somewhat different pathologies; once the Einstein equations are fully solved, many of those specific examples are actually Riemann flat.) 

\noindent
Standard general relativity (the Einstein equations) now yields the stress-energy:
\begin{itemize}
\item 
Eulerian energy density:
\begin{equation}
\label{E:divergence-x}
\rho = {G_{nn}\over8\pi} = 
{1\over16\pi} \left\{ \div \{ \vec v\,K - (\vec v \cdot\grad)  \vec v\}  - {1\over2} \, (\vec\omega \cdot \vec\omega) \right\}.
\end{equation}
Note: (comoving density) = (3-divergence) + (negative semidefinite term). 

\item 
Eulerian flux:
\begin{equation}
\label{E:flux}
 f_i = {G_{ni}\over 8\pi}  =  {1\over8\pi}(K_{ij,j}- K_{,i}) = {1\over 16\pi} \left( \curl (\curl \vec v) \right)_i
\end{equation}
Note: curl of a curl. (So this can sometimes be set to zero.)
\item
Eulerian spatial stresses
\begin{equation}
 \label{E:stress}
\hspace{-20pt}
T_{ij} = {G_{ij}\over 8\pi} =
 {1\over8\pi} \left( \L_n K_{ij}  + K K_{ij} - 2(K^2)_{ij} -  \left(\L_n K  + {1\over2}K^2 + {1\over2} \tr(K^2)\right) \delta_{ij}\right).
\end{equation}
\end{itemize}
The Eulerian density and Eulerian flux are easy, the Eulerian spatial stresses are more complicated.
A key quantity of interest is the average pressure:
\begin{equation}
\bar p = {1\over 3}\,{T_{ij} \,\delta^{ij}}  ={1\over24\pi} \left( -2 \L_n K -{1\over2} K^2 - {3\over2} \tr(K^2)\right).
\end{equation}
This can also be written as:
\begin{equation}
\bar p = {1\over 3}\,{T_{ij} \,\delta^{ij}} = -{1\over24\pi} \left(  2 \nabla_a ( K n^a) -{3\over2} K^2 + {3\over2} \tr(K^2) \right). 
\end{equation}
Equivalently:
\begin{equation}
\bar p= \rho  -{1\over12\pi} \;  \nabla_a ( K n^a). 
\end{equation}
Furthermore, consider the two combinations:
\begin{equation}
\label{E:sec-1}
\rho+3\bar p  = -{1\over4\pi}  \left(  \L_n K  + \tr(K^2) \right),
\end{equation}
\begin{equation}
\label{E:nec-1}
\rho+\bar p = {1\over24\pi} \left( -2 \L_n K + K^2 - 3 \tr(K^2) \right).
\end{equation}
These two quantities are relevant to investigating violations of the SEC and NEC, respectively.~\cite{Santiago:2021-warp}

\section{Forces: Tractor/pressor/stressor beams}

For tractor/pressor/stressor beam spacetimes we now make a specific, concrete choice for the form of the generic warp metric. 
Let us consider a beam pointed in the $\hat z$ direction, and  specialize the flow vector to be of the form:
\begin{align*}
v_x(t,x,y,z) &=  k(t,z)\, x\, h(x^2+y^2),
\\
v_y(t,x,y,z) &=   k(t,z)\, y\, h(x^2+y^2),
\\
v_z(t,x,y,z) &= v(t,z)\, f(x^2+y^2).
\end{align*}
Note the introduction of two envelope functions, $k(t,z)$ and $v(t,z)$; and two profile functions, $h(x^2+y^2)$ and $f(x^2+y^2)$.
Note that for simplicity we have also imposed cylindrical symmetry (in Cartesian coordinates). 

Now consider the forces such a tractor/pressor/stressor beam spacetime would exert on a target with cross-section $\mathbb{S}$ that is at time $t$ placed at position $z$. 
\begin{equation}
\label{Fz:general}
F(t,z) = \pm \int_{\mathbb{S}} T_{zz}(t,x,y,z) \, \d x \, \d y.
\end{equation}
There are two special cases to consider depending on whether the beam is wide or narrow compared to the cross-sectional area $A(\mathbb{S})$ of the tatget.
\begin{itemize}
\item 
Narrow beam:
\begin{equation}
\label{Fztotal}
F(t,z) = \pm \int_{\mathbb{R}^2} T_{zz}(t,x,y,z) \, \d x \, \d y.
\end{equation}
\item
Wide beam:
\begin{equation}
\label{FzA}
F(t,z) = \pm  T_{zz}(t,0,0,z) \, A(\mathbb{S}).
\end{equation}
\end{itemize}
Let us now calculate, setting $u=x^2+y^2$ for conciseness.
\begin{itemize}
\item 
Narrow beam:
\begin{align}
\label{E:critical} 
F(t,z) &= -{1\over2} v(t,z)^2 \int_0^\infty u [f'(u)]^2\d u 
\nonumber\\
	&+ {1\over 8} \int_0^\infty u 
	\left[ v(t,z) f'(u) + {1\over2} \partial_z k(t,z) h(u)\right]^2\d u.
\nonumber
\end{align}	
This is of indefinite sign, depending delicately on the envelope functions and profile functions, potentially allowing either tractor/pressor behaviour. 

\item
Wide beam:
\begin{equation}
\label{E:generic-wide}
        F(t,z) = -{h(0)\over8\pi} \left\{2f(0) \;v(t,z) \partial_z k(t,z)+ 
	2\;\partial_t k(t,z)+3h(0) \; k(t,z)^2 \right\} A(\mathbb{S}).\qquad
\end{equation}

This is of indefinite sign, depending delicately on the envelope functions, potentially allowing either tractor/pressor behaviour. 
\end{itemize}
There are very many special cases to study, depending on one's choice of envelope and profile functions. See reference~\refcite{Santiago:2021-tractor} for extensive details and discussion.

{\small
\begin{table}[h!]
		\centering
\begin{tabular}{| c | c | c | c |}
		\hline
\textbf{Generic Nat\'ario} & \textbf{Modified Alcubierre} & \textbf{Zero Expansion} & \textbf{Zero Vorticity} \\ 
                 \hline
\textbf{envelope} $k(t,z)$ & $0$ & $-\partial_z v$ & $\Phi$ \\ 
                 \hline 
\textbf{envelope} $v(t,z)$ & $v$ & $v$ & $\partial_z \Phi$ \\ 
                 \hline 
\textbf{profile} $h(u)$ & $0$ & $h$ & $2f'$ \\ 
                \hline
\textbf{profile} $f(u)$ & $f$ & $2(h+uh')$ & $f$ \\
               \hline 
\end{tabular}
\end{table}
	}
One can easily tune the resulting forces in a wide variety of ways. In particular we can define a \emph{stressor} beam to be one where the \emph{net} force is zero, but forces alternate between attractive and repulsive as one moves across the cross section of the target, thereby stressing the target in a (in principle) controllable  manner.

\section{Energy conditions --- specifics}

\enlargethispage{40pt}
In this context, all of the classical point-wise energy conditions are violated.
The arguments are fully generic to tractor/pressor/stressor beam spacetimes, 
and, with suitable subsidiary conditions, apply also to warp drive configurations.  (In the warp drive context, many special cases have been well known for over 20 years.)

Generically for either tractor/pressor/stressor or warp drive configurations we have:
\begin{equation}
\label{E:divergence-x2}
\rho = 
 \hbox{(3-divergence)}    - {1\over32\pi} \, (\vec\omega \cdot \vec\omega).
\end{equation}
Then for all tractor/pressor/stressor beams (where by construction all fields fall off sufficiently rapidly at spatial infinity):
\begin{equation}
\int \rho \; \d^3 x = -  {1\over32\pi} \, \int (\vec\omega \cdot \vec\omega) \; \d^3 x \leq 0.
\end{equation}
This kills off the WEC for all tractor/pressor/stressor beams. 
For warp drives one must sometimes be more careful, see the Appendix, and to prove or disprove the WEC one must check both the Eulerian energy density $\rho$ \emph{and} the NEC ($\rho+\bar p\geq 0$).

Generic to either tractor/pressor/stressor or warp drive configurations we have:
\begin{equation}
\bar p= \rho  -{1\over12\pi} \;  \nabla_a ( K n^a). 
\end{equation}
\begin{equation}
\int \bar p \; \d^3 x =  \int \rho \; \d^3 x -  {1\over12\pi}  \int \partial_t K \; \d^3 x.
\end{equation}
\begin{equation}
\int (\rho + \bar p) \; \d^3 x =  -  {1\over16\pi} \, \int (\vec\omega \cdot \vec\omega) \; \d^3 x
 -  {1\over12\pi}  \int \partial_t K \; \d^3 x.
\end{equation}
For any tractor/pressor/stressor/beam configuration (and for any suitably localized warp bubble):
\begin{equation}
\int (\rho + \bar p) \; \d^3 x =  -  {1\over16\pi} \, \int (\vec\omega \cdot \vec\omega) \; \d^3 x
 -  {1\over12\pi}  \partial_t \int  K \; \d^3 x.
\end{equation}

Furthermore, for any tractor/pressor/stressor/beam configuration (and for any suitably localized warp bubble) we also have:
\begin{equation}
  \int K \; \d^3 x = 0 .
\end{equation}

Therefore, for any tractor/pressor/stressor/beam configuration (and for any suitably localized warp bubble) we conclude:
 \begin{equation}
\int (\rho + \bar p) \; \d^4 x =  -  {1\over16\pi} \,  (\vec\omega \cdot \vec\omega) \; \d^4 x
 \leq 0.
 \end{equation}

This kills off the NEC.

That is, all of the classical point-wise energy conditions are violated. This certainly holds for any tractor/pressor/stressor/beam configuration, and for any suitably localized warp bubble, will also work for warp drive spacetimes. An example of a more delocalized warp bubble is considered in the Appendix. 

\section{Conclusions}

\begin{itemize}
\item 
Traversable wormholes, warp drives, and tractor/pressor/stressor beams can all usefully be ``reverse engineered'', debated, analyzed, and studied within the framework of standard general relativity.
\item
Reverse engineering a specified interesting spacetime reduces the task of solving the Einstein equations to a relatively straightforward exercise of differentiation, rather than the more complicated task of integration. 
\item
Traversable wormholes, warp drives, and tractor/pressor/stressor beams all violate all of the standard point-wise energy conditions.
\item 
Even in various modified theories of gravity, wormholes, warp drives, and tractor/pressor/stressor beams
violate the null convergence condition and timelike convergence condition. 
\item
Violation of the classical energy conditions is not an absolute prohibition on ``interesting'' physics --- it is merely an invitation to think very carefully about the underlying physics. 
\end{itemize}

\enlargethispage{20pt}
\appendix{Falloff conditions at spatial infinity}
For tractor/pressor/stressor beams suitable falloff conditions at spatial infinity are inherent to the definition, and utterly automatic. 
Similarly, for Alcubierre warp drives and zero-expansion warp drives suitable falloff conditions at spatial infinity are also inherent to the definition and utterly automatic. For zero vorticity warp drives one must be just a little more careful. 

As an example, consider the general framework 
\begin{equation}
\d s^2 = - \d t^2  + \delta_{ij}\; \left(\d x^i - v^i(t,\vec x) \,\d t\right) \; 
\left(\d x^j  - v^j(t,\vec x) \,\d t\right),
\end{equation}
and for one's warp field take the specific case 
\begin{equation}
\label{E:warp5}
v^i(t,\vec x) = -\sqrt{2m\over r} \; \hat r^i + v^i_\infty(t).
\end{equation}
This corresponds to a warp bubble, centred at the origin, in a coordinate system that is moving with the warp bubble. The presence of the warp velocity term, $v^i_\infty(t)$,  is essential, since otherwise the warp bubble is not moving with respect to spatial infinity, (the ``fixed stars''), which would make the warp field trivial.

For the specific warp field (\ref{E:warp5}) it is easy to 
calculate
\begin{equation}
K = \div \vec v = - {3\over2} {\sqrt{2m}\over r^{3/2}},
\end{equation}
and see that this is independent of the warp velocity $v^i_\infty(t)$.

Then for the Eulerian energy density and flux one has the particularly simple results:
\begin{equation}
\rho =0; \qquad \hbox{and} \qquad \vec f = 0.
\end{equation}
This does \emph{not} mean that the WEC is satisfied since one still has to check $\rho+\bar p$.

Specifically, in this particular warp field we have
\begin{equation}
\rho+\bar p = \bar p =  -{1\over12\pi} \;  \nabla_a ( K \, n^a)
= -{1\over12\pi} \;  \div ( K \, \vec v(t,\vec x)). 
\end{equation}
That is
\begin{equation}
\rho+\bar p  
= -{1\over12\pi} \;  \div \left(   {3\over2} \; {2m\over r^{2}} \; \hat r +  K\vec v_\infty \right)
= -{1\over12\pi} \; \vec v_\infty(t) \cdot \grad K.
\end{equation}
Finally
\begin{equation}
\rho+\bar p  
=  - {3\over 16\pi} {\sqrt{2m}\over r^{5/2}}  \;\; \vec v_\infty(t) \cdot \hat r
\end{equation}
But the factor $\vec v_\infty(t) \cdot \hat r$ explictly changes sign depending on whether one is ``in front of'' or ``behind'' the warp bubble. 
Thus this warp bubble explicitly violates the NEC, and so violates the WEC as well --- this is a specific and explicit example of why merely checking the non-negativity of the Eulerian density is simply insufficient for checking the WEC.

More generally, note that an \emph{asymptotic} version of this argument will still apply to any warp field that is asymptotically of the form (\ref{E:warp5}):
\begin{equation}
\label{E:warp6}
v^i(t,\vec x) = -\sqrt{2m\over r} \; \left\{1+o(1/r)\right\} \; \hat r^i + v^i_\infty(t).
\end{equation}
This now applies to any warp bubble that asymptotes to Schwarzschild spacetime at large distances.

\section*{Acknowledgments}
MV was directly supported by the Marsden Fund, via a grant administered by the Royal Society of New Zealand.\\
JS  acknowledges indirect financial support via the Marsden fund, 
administered by the Royal Society of New Zealand.\\
SS acknowledges support from both the technical and administrative staff at the Charles University, and specific 
financial support via OP RDE project number CZ.02.2.69/0.0/0.0/18\_053/\-0016976 (International mobility of research).


\bigskip
\hrule
\end{document}